# Perovskite topological exciton-polariton disclination laser at room temperature

Feng Jin[1,#], Subhaskar Mandal[1,#], Xutong Wang[1], Baile Zhang[1,2,*], and Rui Su[1,3,*]

[1]Division of Physics and Applied Physics, School of Physical and Mathematical Sciences, Nanyang Technological University, Singapore, Singapore

[2]Centre for Disruptive Photonic Technologies, Nanyang Technological University, Singapore, Singapore

[3]School of Electrical and Electronic Engineering, Nanyang Technological University, Singapore, Singapore

[#]These authors contributed equally to this work.

[*]E-mail: *blzhang@ntu.edu.sg* (B.Z.); *surui@ntu.edu.sg* (R. S.)

**Abstract:**

Topologically nontrivial systems can be protected by band topology in momentum space, as seen in topological insulators and semimetals, or real-space topology, such as in lattice deformations known as topological disclinations (TDs). TDs, with inherent chiral symmetry, can support localized states pinned spectrally to the middle of the topological gap, preventing hybridization with bulk bands, and making them promising for topological lasers. Here, we experimentally realize a $C_{4v}$ symmetric TD laser based on perovskite exciton-polariton lattices at room temperature. Protected by the chiral and point group symmetries of the lattice, the TD state emerges in the middle of the gap and at the core of the perovskite lattice. Under a non-resonant pulsed excitation, coherent polariton lasing occurs precisely at the TD state with a low threshold of 9.5 μJ/cm$^2$, as confirmed by momentum space and real space spectra measurements. This study not only introduces a class of symmetry-protected topological lasers, but also expands the landscape for exploring exciton-polariton light-matter interactions with novel topological structures.

**Introduction**

Topological photonic systems have attracted significant attention due to their remarkable resilience to imperfections[1,2], offering promising prospects for the development of robust devices like topological lasers[3-6]. The robustness of these systems is intricately tied to a bulk topological invariant, typically calculated in the momentum space of periodic structures[7]. For instance, in the one-dimensional Su-Schrieffer-Heeger system, a winding number corresponding to a non-trivial Berry phase within the Brillouin zone leads to the emergence of robust topological modes at the edges of the finite system. However, many physical systems, including quasicrystals[8], grain boundaries[9], and lattice defects[10], exhibit an aperiodic nature, making them incompatible with momentum space topology. Despite this challenge, such systems can still host robust topological modes protected by real-space topology, often influenced by different spatial symmetries.

A prime example of real-space topology is found in topological disclinations (TDs), which are lattice defects arising from the crystallographic disruption of the lattice's periodic nature. Previous works on TDs primarily focused on momentum space topological systems that host topological modes in both the presence and absence of TDs[11-15]. However, these systems inherently break chiral symmetry[14,15], requiring additional fine-tuning to optimize practical applications such as topological lasing[16]. On the other hand, topological modes induced solely by TDs, through the interplay of real-space topology and spatial symmetries, offer a unique advantage[17]. In such cases, the system without the TD exhibits trivial topology in momentum space and cannot host any topological modes. Real-space topology-protected TDs intrinsically respect chiral symmetry, placing the topological modes in the middle of the bulk bandgap without the need for fine-tuning, making them a promising platform for realizing topological mid-gap states.

Another key advantage of TDs compared to periodic lattices is their ability to facilitate the appearance of topological states within the lattice bulk, reducing sensitivity to boundaries. Consequently, TD modes exhibit enhanced robustness compared to systems with typical zero-dimensional topological modes localized at edges, which face challenges in maintaining pristine edges during the nano-fabrication process. Thus, while the mid-gapness of TD modes provides spectral isolation, their bulk localization ensures maximal confinement and enhanced robustness against edge defects, essential for realizing robust high-purity single-frequency topological lasers. Despite tremendous experimental advances in lasing within topological states since 2018, employing various schemes and mechanisms[18-29], the realization of real-space topology-protected mid-gap state TD lasing remains elusive.

Among various systems, exciton-polaritons resulting from the strong coupling of excitons to microcavity photons, have emerged as a promising platform for topological lasers[3,30-34]. Their unique capability to spontaneously condense into collective condensates, allows coherent lasing without the need of population inversion, offering the potential for a lower power threshold compared to conventional lasers[35,36]. While experimental evidence of topological polariton lasing has been successfully observed in GaAs microcavities operating at cryogenic temperatures[3,30], recent advances in novel semiconductor emitters, such as inorganic perovskites and organic fluorescent proteins, have further propelled the development of topological polariton lasers into the domain of room temperature operation, including 1D SSH lasers and two-dimensional (2D) higher-order topological lasers[31-34]. Despite these breakthroughs, earlier demonstrations on topological polaritons have been limited to periodic lattices, while the exploration of aperiodic structures has been largely uncharted as of now.

In this study, we present the experimental realization of single-mode topological polariton lasing at mid-gap TD state in a 2D aperiodic perovskite lattice at room temperature. By employing a $C_6$ symmetric topological crystalline insulator (TCI) lattice model and introducing a 120° Frank angle, we generate a $C_4$ symmetric disclination lattice using the Volterra process. Protected by the chiral symmetry and point group symmetry of the lattice, the TD state could pin to the middle of the topological bandgap spectrally and the core of the lattice spatially. Through the angle-resolved photoluminescence measurements, we experimentally validate the emergence of mid-gap TD state both in momentum and real spaces. Under a pulsed excitation, we achieve single-mode polariton lasing exactly at mid-gap TD state at room temperature, with a lasing threshold of 9.5 $\mu J/cm^2$ and a temporal coherence of 1.15 ps.

Results

Design of the TD lattice

In this work, we realize the TD lattice by etching the spacer layer of the $CsPbBr_3$ perovskite planar microcavity, as shown in Figure 1a (See methods for more details). It consists of a bottom distributed bragg reflector (DBR), an all-inorganic perovskite ($CsPbBr_3$) nanoplatelet, a spacer lattice pattern and a top DBR. Similar method has been employed to achieve polariton condensation and topological polariton edge states in perovskite periodic lattices at room temperature[31,33,37]. In order to generate the TD lattice structure, we begin with a $C_6$ symmetric TCI, where each unit cell consists of six sites, as depicted in cyan in Figure 1b. The $C_4$ symmetric disclination lattice is obtained by removing a 120° Frank angle sector from the $C_6$ symmetric lattice and then joining them via the Volterra process. This process forms a TD at the center of the lattice, resulting in a central unit cell with four sites, while the remaining unit cells retain six sites. Depending on the interplay between intracell ($t_1$) and intercell ($t_2$) couplings, different scenarios can be achieved in the same lattice structure. We initiate our investigation by theoretically examining the lattice within the framework of tight-binding (TB) description (see Supplementary Section 1). In the scenario where $t_1 < t_2$, the system exhibits an obstructed atomic limit (OAL) topological phase, wherein the Wannier centers are positioned at the edges of the unit cell[13]. In this regime, the system can be characterized by a non-trivial second-order topological index, and the eigenenergies of the system reveal four fractionally charged zero-energy modes within the bulk bandgap, spatially localized at the corners of the lattice. In the scenario where $t_1 > t_2$, the three Wannier centers are positioned at the center of the unit cell resulting in the failure of the Wanier type topology and the absence of the zero-energy corner modes[13]. However, the system maintains another OAL topological phase and gives rise to two zero-energy TD modes pinned to the core of the lattice (see Supplementary Section 1). Unlike the zero-energy corner modes in the previous case, the TD modes do not exhibit fractionalized charge. Instead, their topological protection relies on the symmetries of the lattice in the form of the chiral symmetry of the entire lattice and the $C_{4v}$ ($C_4$ + reflection symmetry) point group symmetry of the disclination lattice[17,38]. This protection manifests in two ways: firstly, the $C_{4v}$ symmetry establishes a 2D irreducible representation, where two combinations of the TD modes become rotationally symmetric partners with identical energy and thus result in degeneracy. Secondly, the chiral symmetry compels the mode energies to be equal and opposite. However, it conflicts with the degeneracy imposed by $C_{4v}$ symmetry, forcing the TD modes to be pinned at the middle of the topological gap with zero energy and ensuring both spectral isolation and maximal confinement (see Supplementary Section 4).

To realize the TD modes in exciton-polaritonic lattices, we control the intracell and intercell couplings by adjusting the distances between micropillars with a diameter of 650 nm. As shown in Figure 1c, the lattice possesses $C_{4v}$ symmetry about the core of the TD lattice, with intracell coupling enhanced by strong overlap between adjacent pillars, compared to the weaker intercell coupling where pillars do not overlap. We further employ the molecular dynamics simulations to extract the coordinates of each micropillar for the fabrication of the TD lattice[39]. By diagonalizing the continuous Schrödinger equation Hamiltonian, we obtain the polariton eigenenergies and achieve a topological gap of approximately 27 meV, with two topologically protected TD modes sitting in the middle of the bandgap spectrally (Figure 1d) and spatially in the core of the lattices (Figure 1e), which is in good agreement with the TB results.

**Characterization of the bulk and topological mid-gap states in the TD lattice**

In our TD lattice design, the combination of chiral symmetry and the $C_{4v}$ ($C_4$ + reflection symmetry) point group symmetry allows to pin the TD states to the exact middle of the gap. However, during the actual lattice fabrication, the introduction of the onsite energy variations could disrupt the chiral symmetry[3,40], shifting the TD modes towards the bulk band edges. In our experiment, we optimize the lattice fabrication process to achieve homogenous lattice by employing atomically-flat perovskite samples and layer-by-layer lattice etching process (Methods), thereby minimizing onsite energy variations. Figure 2a displays the typical atomic force microscopic (AFM) image of our TD lattice, where the lattice sites are clearly distinguished from the magnified image (left panel in Figure 2b). To further assess the thickness uniformity, we collect the height profile along the white dashed line indicated in Figure 2b (left), revealing the homogeneous sample thickness ~ 85 nm at the lattice sites (right panel in Figure 2b). To experimentally demonstrate the emergence of TD state in the polariton lattice, we characterize the band structure through mapping the angle-resolved photoluminescence spectra (Methods). The entire 2D polariton lattice with a detuning of ~ -180 meV is non-resonantly excited using a continuous-wave laser at 2.713 eV with a pumping spot diameter of 25 μm at room temperature. By collecting the emission from the bulk area and the disclination area, framed by the white and red dashed lines in Figure 2a, respectively, we observe distinct scenarios. Figure 2c illustrates the angle-resolved photoluminescence spectrum from the bulk area along the $k_x$ direction at $k_y$ = 0. We observe multiple discrete states from the bulk area and they originate from the coupling of *s* orbital modes of polaritons inside pillars, owing to the small pillar size. Inside the spectrum, we also observe a huge bandgap of ~ 35 meV without any states inside at around 2.26 eV. As shown in Figure 2d, the emission from the disclination area exhibits very similar behaviors, but notably an additional discrete state appears inside the topological gap at around 2.26 eV, which corresponds to the TD state and aligns well with our theoretical calculation (Figure 1d and Supplementary Section 6). Benefitting from the small pillar size and TD lattice design, the topological gap can open more than 30 meV which is at least three times larger than previous demonstrations in polariton systems[3,30-34], enabling the generation of topological state with significant spectral isolation. We further compare the polariton photoluminescence spectra at $k_x$ = -4.9 μm$^{-1}$ (black dashed lines in Figure 2c, d) by fitting with Gaussian functions, evidencing the emergence of TD state inside the gap. To enhance the understanding, we measure the real space profile at the TD state (E = 2.259 eV), indicated by red arrow in Figure 2e. As shown in Figure 2f, the emission from the TD state strongly localizes at the four sites in the core of the lattice, aligning well with the theoretical calculation (Figure 1e). These results collectively prove the emergence of TD state which spectrally sit in the middle of the gap and spatially locate at the core of the lattice.

**Exciton-polariton lasing at the TD state**

Compared with the conventional topological lasers that rely on population inversion, one of the significant advantages of exciton-polaritons is their ability to realize non-equilibrium condensation at high temperatures with low power consumption, accompanied by laser-like coherent emission. The driven-dissipative nature of polaritons allows them to condense into non-ground states[3,30-34,37,41], providing the possibility of condensation at the TD state. Here, we employ a lattice with a detuning of ~ -120 meV for better relaxation into the TD state and pump the lattice non-resonantly with a pulsed excitation at 3.1 eV (Methods). As shown in Figure 3a, under a low excitation power of 0.6 $P_{th}$ ($P_{th}$ represents the threshold pump fluence), the dispersion of the TD lattice exhibits similar features as the linear region in Figure 2d, and its corresponding real-space image indicates a nearly uniform density of polaritons throughout the pumping area (Supplementary Section 7). Subsequently, as the excitation power increases, polaritons tend to condense at the TD state at $P = P_{th}$ (Figure 3b) with more occupations than other states. This phenomenon is further elucidated by the real-space image, illustrating a large fraction of polaritons accumulating at the core of the TD lattice (Supplementary Section 7). While beyond the threshold at 2.0 $P_{th}$, the emission from the TD state totally dominates both spectrally and spatially over other states (Figure 3c), suggesting the realization of polariton condensation at the TD state. To further characterize this transition, Figure 3d presents the polariton emission spectra at $k_x$ = -5.2 μm$^{-1}$ under different excitation powers, including $P$ = 0.6 $P_{th}$, 1.0 $P_{th}$, 1.3 $P_{th}$ and 2.0 $P_{th}$. The intensities of the spectra under $P$ = 0.6 $P_{th}$, 1.0 $P_{th}$ are multiplied by a factor of 5 for better comparison. It is evident that polariton condensation occurs in our lattice system, transitioning from multiple states emission below the threshold to single-mode lasing above the threshold. In order to quantify the transition, we present the evolution of emission intensity, linewidth, and peak energy extracted from the TD state as a function of pumping fluence. As depicted in Figure 3e, the integrated emission intensity curve displays an *S*-shape behavior, nonlinearly increasing by three orders of magnitude beyond the threshold of $P_{th}$ = 9.5 μJ/cm². Moreover, the linewidth of the TD state rapidly decreases from 6.7 to 2.1 meV at the threshold. Simultaneously, the peak energy of the TD state continuously blueshift due to the repulsive interactions in polariton system. In the meantime, we measure the linear polarization of the TD state lasing under $P = 2.0\, P_{th}$ and it exhibits a highly vertical linear polarization up to 80% along the perovskite crystal axis (Supplementary Section 8).

One of the key features of polariton condensation is the build-up of spontaneous coherence, leading to coherent emission from the condensate[36]. We further characterize the coherence from the TD state above the threshold with a home-built Michelson interferometer. One arm is replaced with a retroreflector to revert the real space image in a centrosymmetric way. Another arm is placed onto a moving stage for investigating the coherence in the time domain, as illustrated in Figure 4a. The interference fringe contrast thus serves as a measurement of the phase coherence between points located at *r* and *-r* respect to the center. By changing the mirror position with the moving stage, one can continuously modify the time delay between the two arms and monitor the evolution of interference fringe contrast. Figure 4b shows the superimposed real-space image from the TD state at a time delay $\Delta\tau$ = 0 ps under $P = 2.0\, P_{th}$. Clearest interference fringes can be observed throughout the whole core area of the lattice (Figure 4b inset), suggesting the buildup of phase coherence. Increasing the time delay, for example to 0.73 ps and 1.80 ps in Figure 4c, the interference fringe contrast becomes weaker and weaker. We further extract the visibility of the interference pattern under different time delays, which is defined as $V = (I_{max} - I_{min})/(I_{max} + I_{min})$ and $I_{max\,(min)}$ is the maximum (minimum) intensity of the fringes. As shown in Figure 4d, the interference fringe contrast exhibits exponential decay as a function of the time delay. By fitting it with an exponential function, we extract a coherence time of 1.15 ps for polariton lasing at the TD state.

**Discussion**

In conclusion, we have demonstrated topological polariton TD state lasing in 2D aperiodic perovskite lattices. Protected by $C_{4v}$ symmetry and chiral symmetry, the TD modes pin to the middle of the topological gap spectrally and the core of the lattice spatially, exhibiting robustness against hybridization with the bulk. Under a non-resonant pulsed excitation, polaritons tends to condense exactly at the TD state, leading to single-mode coherent lasing with a low threshold of 9.5 μJ/cm² and a coherence time of 1.15 ps. By harnessing the unique properties of exciton-polariton systems, our study not only introduces a novel class of topological lasers, but also lays the groundwork for exploring polaritonic devices based on nonlinear effects within TDs. Our research paves the way for advancing both fundamental understanding and practical utilization of topological defects in optics and photonics, opening new avenues for innovative applications and discoveries in this rapidly evolving field.

**Methods**

**Perovskite TD lattice fabrication**

The perovskite microcavity is composed of a bottom DBR, an all-inorganic perovskite (CsPbBr₃) nanoplatelet, a spacer lattice pattern and a top DBR. In detail, the bottom DBR consists of 15.5 pairs of TiO$_2$ (62.7nm) and SiO$_2$ (92.1nm), which is fabricated by an electron beam evaporator (Cello 50D). The single-crystal CsPbBr$_3$ is grown on mica via chemical vapor deposition method as described before[42]. The 65 nm-thick all-inorganic perovskite nanoplatelet is then transferred onto the bottom DBR by using Scotch tape. Next, we spin-coat an 85 nm-thick ZEP520A layer as electron beam resist onto the sample, and then expose it layer by layer using electron beam lithography to ensure the lattice structure has homogeneous thickness[37]. Lastly, the top DBR consisting of 8.5 pairs of TiO$_2$ (62.7nm) and SiO$_2$ (92.1nm) are deposited on top by the electron beam evaporator to complete the whole fabrication process.

**Optical spectroscopy characterizations**

The energy-resolved momentum-space and real-space photoluminescence characterizations are detected by a home-built angle-resolved spectroscopy setup with Fourier optics. The emission signal of the polaritons from the perovskite microcavity can be detected by measuring their photon part leaking out of the microcavity. Then the emission is collected by a 50× objective lens (NA = 0.75) and sent to a 550 mm focal length spectrometer (Horiba iHR550) with a grating (600 lines/mm) and a liquid nitrogen cooled CCD (256 ×1,024 pixels). In the linear region, the perovskite lattice is pumped by a continuous wave laser (457 nm) with a pumping spot of ~25 μm. In the nonlinear region, the lattice is non-resonantly pumped by a pulsed laser (wavelength: 400 nm, pulse duration: 100 fs and repetition rate: 1kHz) with a homogeneous pump spot of ~13 μm. The real space emission images of the polariton condensate can also be sent to a home-built Michelson interferometer to obtain the time-dependent interference images, where one arm is replaced with a retroreflector to revert the image in a centrosymmetric way.

**Theoretical calculations**

The tight-binding results are obtained numerically using custom code developed in MATLAB. For the continuous modeling of the system, we employ coupled Schrödinger equations (CSEs) under the mean-field approximation. The CSEs are expressed as:

$$i\hbar\frac{\partial \psi_p(x,y,t)}{\partial t} = \left[-\frac{\hbar^2 \nabla^2}{2m_p} + V(x,y)\right]\psi_p(x,y,t) + \frac{g_0}{2}\psi_e(x,y,t),$$

$$i\hbar\frac{\partial \psi_e(x,y,t)}{\partial t} = E_{ex}\psi_e(x,y,t) + \frac{g_0}{2}\psi_p(x,y,t).$$

Here, $\psi_p$ and $\psi_e$ represent the mean-field wave functions of photons and excitons, respectively. $m_p = 0.85 \times 10^{-5} m_e$ denotes the isotropic cavity photon mass, with $m_e$ representing the free electron mass. $V(x,y)$ represents the in-plane photon confinement potential, as depicted in Figure 1c, while $g_0$ denotes the Rabi splitting. Due to the heavy exciton mass, excitons are treated as dispersion-less, having energy $E_{ex}$. The Laplacian $\nabla^2$ is expressed using the fourth-order finite difference method. The Hamiltonian corresponding to the CSEs is then diagonalized to determine the eigenenergies and eigenstates of the system, as illustrated in Figure 1d,e. Other parameters include $g_0$=120 meV and $E_{ex}$=2.407 eV.

**Acknowledgements**

R.S. gratefully acknowledge funding support from the Singapore Ministry of Education via the AcRF Tier 2 grant (MOE-T2EP50222-0008) and Tier 1 grant (RG80/23). R.S. also gratefully acknowledges funding support from the Nanyang Technological University via a Nanyang Assistant Professorship start-up grant. R.S. and B.L. Zhang gratefully acknowledge funding support from the Singapore National Research Foundation via a Competitive Research Program (grant no. NRF-CRP23-2019-0007).

**Author contributions**

R.S. and B.L.Z supervised the project. F.J. fabricated the devices and conducted all the experiments. S.M. conceived the lattice model and performed the theoretical calculations with input from B.L.Z. X. T. W. provided help on the optical measurements. F.J., S.M., R.S. and B.Z. collectively wrote the manuscript with input from all authors.

**Competing interests**

The authors declare that they don't have competing interests.

**Data availability**

All experimental data that support the plots within this paper are available from the corresponding author upon reasonable request.

**Code availability**

The codes are available from the corresponding author upon reasonable request.

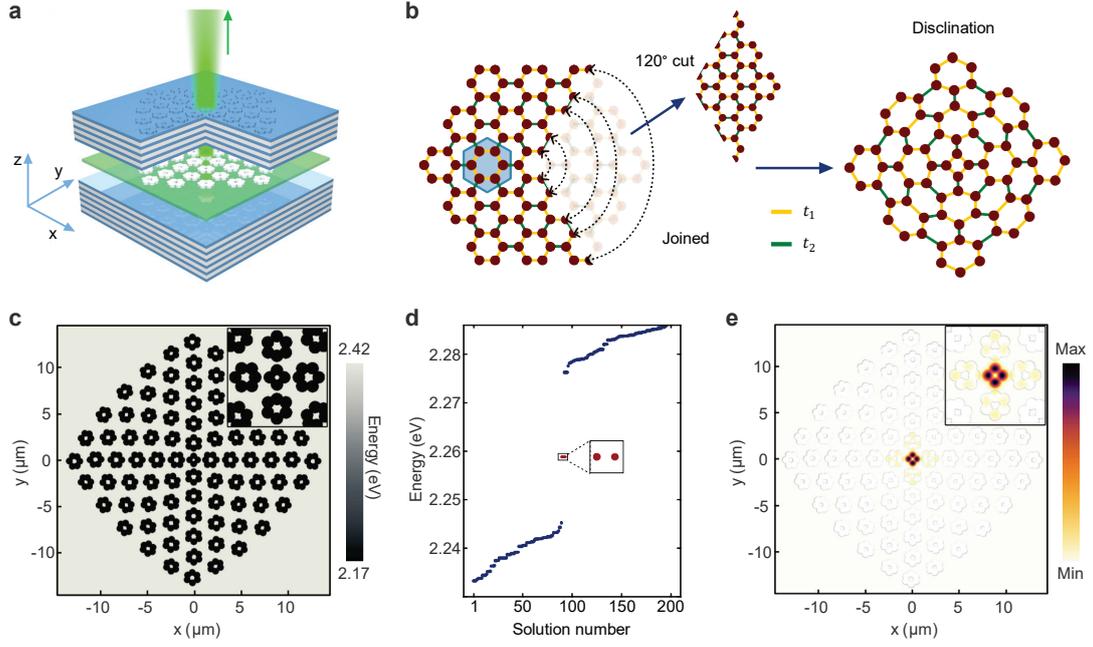

**Fig. 1 Scheme and Design of TD Laser. a,** Schematic illustration of an exciton-polariton topological disclination laser. **b,** Generation of a $C_4$ symmetric disclination lattice structure from an undeformed $C_6$ symmetric TCI through the cut-and-join Volterra process. Here, $t_1$ and $t_2$ represent the intra and intercell couplings, respectively. **c,** Potential landscape for cavity photons in the XY plane. **d,** Eigenenergies of the polaritons corresponding to the disclination lattice shown in **c**, revealing two TD modes located at the middle of the bulk bandgap. **e,** Spatial profile of the mid-gap TD modes demonstrating localization at the core of the TD lattice. For clarity, the potential profile is superimposed with the eigenmode profile.

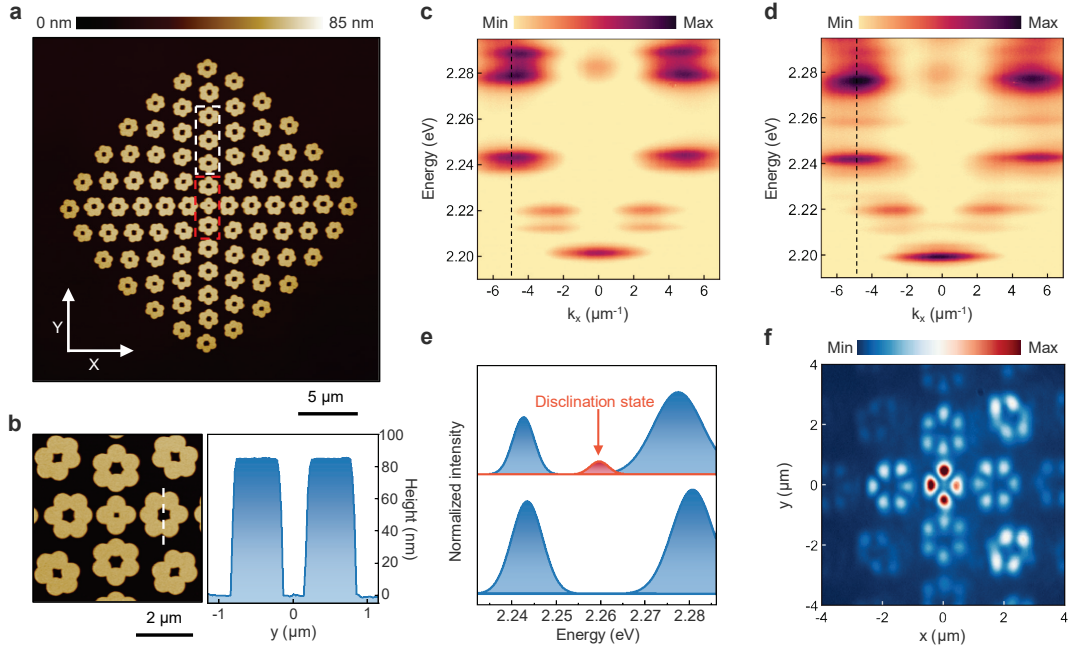

**Fig. 2 Experimental demonstrations of TD state in the perovskite lattice. a,** Atomic force microscopy image of the 2D TD lattice with a thickness around 85 nm on perovskite layer before the deposition of top DBR. The white dashed lines and red dashed lines represent the emission collection area in **c** and **d**, respectively. **b,** Magnified atomic force microscopy image of the TD lattice core (left) clearly illustrating the lattice sites. The height profile (right) of the perovskite TD lattice along the white dashed line in the left image, showing the excellent homogeneity of the TD lattice. **c,** Momentum-space polariton energy dispersions of bulk area at $k_y = 0$ μm$^{-1}$. The black dashed line is at $k_x = -4.9$ μm$^{-1}$. **d,** Momentum-space polariton energy dispersions of topological area at $k_y = 0$ μm$^{-1}$. The black dashed line is at $k_x = -4.9$ μm$^{-1}$. **e,** Polariton emission spectrum at $k_x = -4.9$ μm$^{-1}$ from **c** (bottom) and **d** (top) fitted with Gaussian function, showing the large bulk bandgap of 35 meV and the emergence of the topological disclination state in the middle of the topological gap, indicated by the red arrow. **f,** Experimental real-space image of the perovskite TD lattice at energies of 2.259 eV, indicated by red arrow in **e**.

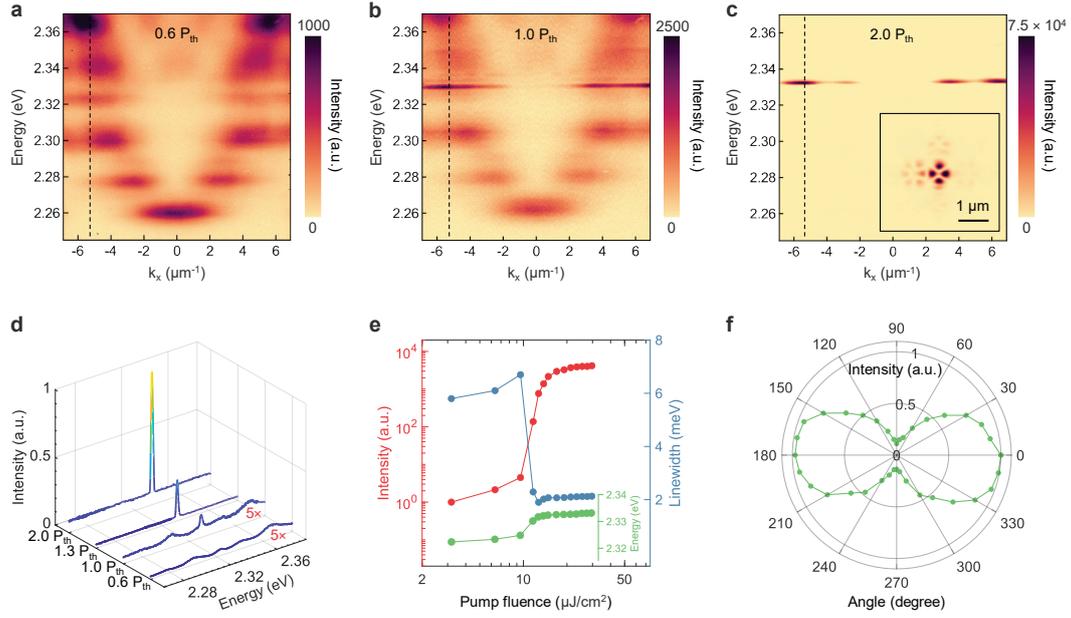

**Fig. 3 Characterization of exciton polariton condensation into the TD state at room temperature.** **a** to **c,** Momentum-space polariton dispersions of the polariton TD lattice with the increase of the excitation power at $P = 0.6\ P_{th}$ (**a**), $P = 1.0\ P_{th}$ (**b**), $P = 2.0\ P_{th}$ (**c**), respectively. The dashed black line is at $k_x$ = -5.2 μm$^{-1}$. Inset of **c** shows the experimental real-space image of polariton lasing at TD state at $P = 2.0\ P_{th}$. **d**. Polariton emission spectrum at $k_x$ = -5.2 μm$^{-1}$ under different excitation power $P = 0.6\ P_{th}$, $1.0\ P_{th}$, $1.3\ P_{th}$, $2.0\ P_{th}$, respectively. The intensities of the spectra under $P = 0.6\ P_{th}$, $1.0\ P_{th}$ are multiplied by a factor of 5 for better comparison. **e**. The evolution of integrated intensity and linewidth of the TD state as a function of pump fluence, showing the superlinear increase of intensity by three orders, substantial narrowing of linewidth at a threshold of $P_{th}$ = 9.5 μJ/cm$^2$. Inset of **e** is the emission energy evolution of the TD state, showing a continuous blueshift. **f**. The TD state lasing intensity as a function of detection polarization angle at the excitation power $P = 2.0\ P_{th}$. Radial axis is normalized intensity.
13

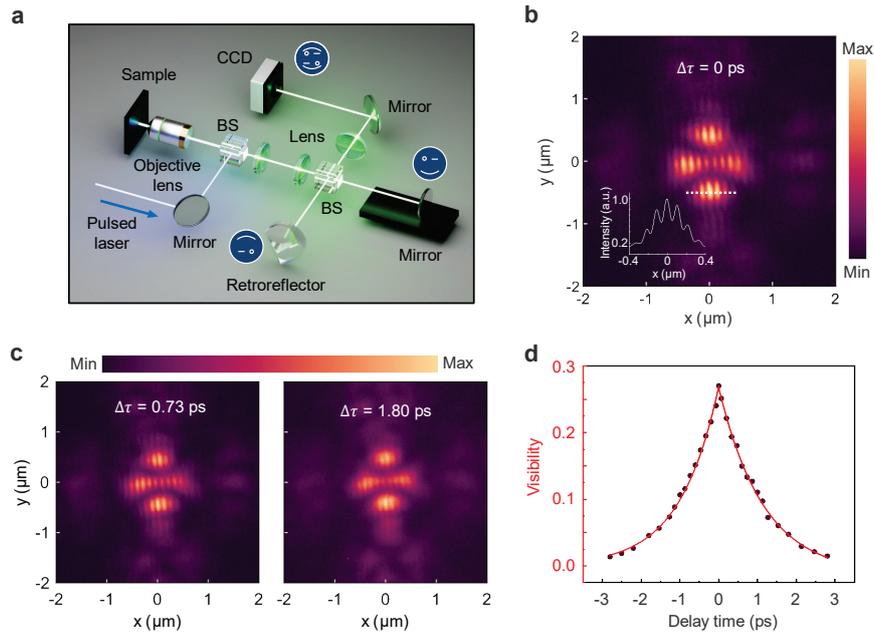

**Fig. 4 First-order coherence measurement of the TD state lasing at room temperature. a,** Schematic of the Michelson interferometer, containing a retroreflector in the reference arm (fixed position) and a flat mirror in the delay arm (mounted on a motorized translation stage). BS represents beamsplitter and CCD represents charged camera device. **b,** Corresponding interference fringe image of the TD state lasing at zero delay time between reference and delay arms. The inset of **b** shows the interference spectra to the white dashed line. **c,** Interference fringe images of TD state lasing at two different delay times between reference and delay arms. Left: delay time equals to 0.73 ps. Right: delay time equals to 1.80 ps. **d,** The visibility of the interference pattern at different delay times, which shows a coherence time of ~1.15 ps.